\documentclass[9pt,conference]{IEEEtran}
\usepackage{dcase2025,amsmath,graphicx,url,times,booktabs, tabularx}

\usepackage{caption}
\usepackage{graphicx}
\usepackage{booktabs}
\usepackage{enumitem, color, slashed, multirow, multicol}
\usepackage[dvipsnames]{xcolor}

\newcommand{\cmt}[1]{\ignorespaces}
\title{Latent Multi-view Learning for Robust Environmental Sound Representations}

\name{Sivan Ding$^{1}$, Julia Wilkins$^{1}$, Magdalena Fuentes$^{1}$, Juan Pablo Bello$^{1}$}
\address{$^1$ Music and Audio Research Laboratory, New York University, New York, NY, USA
}

\begin{document}

\maketitle
\begin{abstract}

Self-supervised learning (SSL) approaches, such as contrastive and generative methods, have advanced environmental sound representation learning using unlabeled data. However, how these approaches can complement each other within a unified framework remains relatively underexplored. In this work, we propose a multi-view learning framework that integrates contrastive principles into a generative pipeline to capture sound source and device information.
Our method encodes compressed audio latents into view-specific and view-common subspaces, guided by two self-supervised objectives: 
contrastive learning for targeted information flow between subspaces, and reconstruction for overall information preservation.
We evaluate our method on an urban sound sensor network dataset for sound source and sensor classification, demonstrating improved downstream performance over traditional SSL techniques. Additionally, we investigate the model's potential to disentangle environmental sound attributes within the structured latent space under varied training configurations.
\end{abstract}


\section{Introduction}

\cmt{As a crucial field in machine listening and scene analysis, environmental sound representation learning plays a key role in tasks such as environmental sound classification and acoustic scene classification.}
Environmental sound representation learning is a crucial field in machine listening, serving as a foundation for tasks like environmental sound classification and acoustic scene analysis.
To address the scarcity of annotations in environmental sound data, self-supervised learning (SSL) aims to approach and even sometimes surpass the performance of supervised methods through self-supervision techniques, such as contrastive and generative objectives.

Contrastive learning (CL) for audio representations exploits the assumption of commonalities across different views of data with the same semantics to create positive pairs, such as using augmented ``views'' of the same audio clip \cite{cola, byol, al2021clar}. These principles have been widely employed for environmental sound \cite{nasiri2021soundclrcontrastivelearningrepresentations, MAHYUB2024SLS, Perera2022adv}.
Generative learning focuses on reconstruction-based objectives to learn an embedding space for a single ``view'' of complete or masked input\cite{kumar2023highfidelityaudiocompressionimproved, défossez2022highfidelityneuralaudio, gong2022ssast, chong2023maskedspec, huang2022masked, beats}. In acoustic scene classification and sound event detection \cite{abesser2017autoencode, zinemanas2021prototype, Abesser2017dimensionality, Wilkinghoff2019opensetauto}, these methods have proven effective at capturing a diverse acoustic content by encoding information across both temporal and feature dimensions.
At the same time, in other domains, SSL research has begun to explore the complementary effects of combining reconstructive and contrastive learning frameworks in various ways. For example, \cite{xu2021multi} and \cite{lee2020private} propose multi-view learning frameworks for visual clustering or classification. These methods simultaneously learn a ``shared'' latent subspace that is invariant across views and a ``private" latent subspace that varies across views, by optimizing view reconstruction from both subspaces within an autoencoder architecture. This approach 
employs a view-contrastive strategy by learning both view-common and view-specific subspaces, without relying on explicit contrastive objectives.
In \cite{gong2023contrastive}, masked reconstruction is combined with CL objectives
in an audio-visual context, but with one \cmt{view-common} shared latent space for each modality view.

In the \cmt{audio}music domain, for pitch-timbre disentanglement, \cite{tanakaunsup} leverages random perturbations to form view-contrasting training strategies within a reconstruction-based pipeline. Similarly, \cite{luo2020unsupervised} studies how pitch-shifting can be used to create auxiliary objectives such as a contrastive loss in addition to single-view reconstruction.
Furthermore, our previous work \cite{wilkins2024selfsupervisedmultiviewlearningdisentangled} uses a multi-view learning framework to learn disentangled latent subspaces for pitch and timbre without explicit contrastive objectives.
These methods provide insights on how we can combine different SSL strategies and potentially disentangle inherent factors in audio. However, such methods haven't been explored in a more challenging context such as real-world environmental sound recordings, and there lacks a thorough investigation into how objective design choices affect latent subspace structures.


In this work, we combine contrastive and generative objectives within a multi-view learning framework to explore their combined impact on environmental sound representations.
Using DAC \cite{kumar2023highfidelityaudiocompressionimproved} as a latent feature extractor, the framework encodes the audio feature into separate private and shared subspaces based on a metadata-driven data pairing strategy, eliminating the need for explicit class annotations. To strengthen the training signal of contrasting views in our multi-view backbone, we investigate similarity or separation-based contrastive objectives on the subspaces between views.
Our evaluation is conducted on an urban sound sensor network dataset (SONYC-UST-V2 \cite{bello2019sonyc}), with performance measured on sound source and sensor classification tasks. In addition, we examine the model's ability to disentangle sound attributes, \cmt{through a subspace classification-based proxy metric,} offering insights into the structure of the learned latent space.
Our contributions are summarized as follows:
\begin{itemize}
    \item{We propose a novel latent multi-view contrastive learning framework for environmental sound representation learning.}
    \item{By simply creating pairs of data with available metadata and training a lightweight autoencoder in a self-supervised manner, we demonstrate a downstream performance boost on sound and sensor classification compared to traditional SSL baselines, achieving results comparable to supervised baselines.}
    \item{We investigate how different combinations of SSL training strategies influence the latent subspace structures, providing insights into environmental sound attribute disentanglement.}
\end{itemize}

\section{Method}





\begin{figure*}
\begin{center}
    \vspace{-0.5cm}
  \includegraphics[width=0.8\linewidth, trim={1cm 0.5cm 1cm 0cm}]{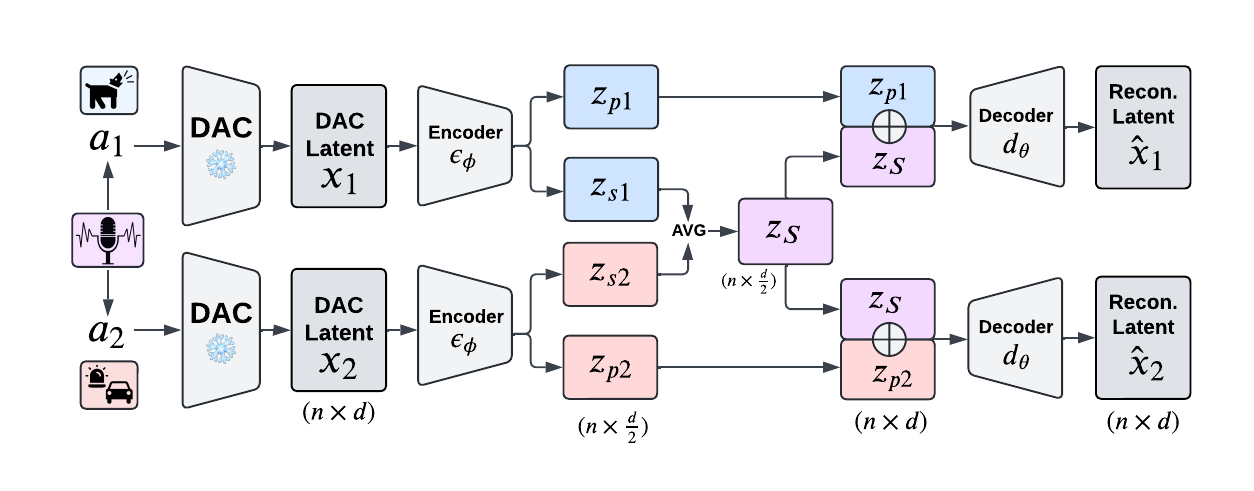}
  \captionsetup{type=figure}
  \captionof{figure}{
     Our self-supervised multi-view framework for environmental sound representation learning. Learned latent subspaces are used for downstream sound source and recording sensor classification tasks.
  }
  \label{fig:block_diag}
\end{center}
\end{figure*}

\subsection{Multi-view Representation Learning}
\label{subsection:MVRL} 
Our proposed method is shown in Figure \ref{fig:block_diag}.
The ``views'' of data used in our method are two audio clips, denoted $\{a_1, a_2\}$. Our system requires a single assumption between these audio clips: that they share a common attribute (e.g., recordings from the same sensor).
We first pass $a_1$ and $a_2$ through a pretrained encoder to obtain 2D embeddings $\{x_1, x_2\}$ of dimension $(n \times d)$, where $n$ is the number of frames and $d$ is the feature dimension.
In this work we use DAC \cite{kumar2023highfidelityaudiocompressionimproved} as the pretrained encoder. 
The DAC latent space, capable of preserving detailed information in environmental sound while reconstructing high-fidelity signals, provides us with a strong foundation for discriminative downstream tasks.

We pass embeddings $x_1$ and $x_2$ through a simple MLP encoder, denoted $\epsilon_\phi$. The weights of $\epsilon_\phi$ are shared across views of data. This encoder projects each input embedding into 
two separate \textit{private} and \textit{shared} latent subspaces, as defined in traditional multi-view learning frameworks \cite{lee2020private}, denoted $z_{pi}$ and $z_{si}$ respectively, where $i$ is the data view (for $i \in 1,2$ in this study). The private subspace is designed to capture information specific to each individual view, while the shared subspace should capture information shared across both views. Latent subspaces $z_{pi}$ and $z_{si}$ of dimension $(n, \frac{d}{2})$ are then averaged to form a joint shared embedding: $z_S$. Additional training strategies can then be applied to these latent subspaces to incentivize desired information flow, 
which we introduce in Section \ref{subsection:designfactor}. Lastly, an MLP decoder $d_{\theta}$ takes the concatenation of $z_{pi}$ and $z_S$ as input and projects back to the original dimensionality,  reconstructing the original pretrained latents: $\{\hat{x_1}, \hat{x_2}\}$.

Importantly, note that because the weights of our projection encoder and decoder are shared across data streams, at inference time, our model can operate using only a single audio input;
pairs are not needed.  The model 
encodes the input audio into the previously learned private, shared, and combined subspaces, yielding $z_p, z_s$ and their concatenation ($z_p \oplus z_s$), reusable for further downstream tasks.




\subsection{Self-supervised Training Strategies}
\label{subsection:designfactor}
To combine generative and contrastive SSL frameworks to learn robust environmental sound representations, we design a suite of training strategies to apply within the multi-view learning backbone.

\textbf{Reconstruction}:
\label{subsection:reconloss}
The base version of our model is trained using a mean squared error reconstruction loss $\mathcal{L}_{rec}$ between the original and reconstructed embeddings, $x_i$ and $\hat{x}_{i}$ respectively, where $i$ is the view index in the data pair and $j$ is the sample in a dataset of size $N$ samples:
\begin{equation}
\small
\label{eq:rec}
\centering
\mathcal{L}_{rec}=\sum_{i} \Biggl[  \frac{1}{N} \sum_{j=1}^{N} \left( x_{i,j} - \hat{x}_{i,j} \right)^2 \Biggr]
\end{equation}   
\textbf{Cosine Distance}: In addition to the base reconstruction loss, we utilize contrastive learning in our framework via objectives based on cosine distance, in which similar embeddings are incentivized to be close together in the latent space and dissimilar embeddings are pushed farther apart \cite{wu2022wav2clip, radford2021learning, guinot2024leaveoneequivariantalleviatinginvariancerelatedinformation}. The general form of this loss term is referred to as $\mathcal{L}_{cos}$. To encourage private latents to capture view-specific information, we introduce a loss term that enforces separation between the two private latents, $z_{p1}$ and $z_{p2}$, by minimizing cosine similarity, denoted $\mathcal{L}_{cos^{-}}$. Along the same lines, 
we maximize cosine similarity between $z_{s1}$ and $z_{s2}$ to encourage the shared latents to contain similar information ($\mathcal{L}_{cos^{+}}$). 

We experiment with these similarity or separation-based configurations on both a batch and sample level. 
The batch-level version includes inherent negatives from other samples in the batch similar to the traditional InfoNCE \cite{oord2019representationlearningcontrastivepredictive} setting, while the sample-level version treats cosine similarity as a simple binary classification, without cross-batch negatives.
The sample-level \cmt{($\mathcal{L}_{cos^{S}}$)} similarity and separation-based objectives are defined below, where $j$ or $k$ refers to the index of samples within a batch of size $B$:
\begin{equation}
\small
\label{eq:cos+samp}
    \mathcal{L}_{cos^{+}}=-\mathbb{E}_{j\in[1,B]}\big [\log \textit{sim}(z_{s1}^i, z_{s2}^i)\big ]
\end{equation}
\begin{equation}
\small
\label{eq:cos-samp}
    \mathcal{L}_{cos^{-}}=-\mathbb{E}_{j\in[1,B]}\big[\log (1-\textit{sim}(z_{p1}^j, z_{p2}^j))\big]
\end{equation}
Similarly, the batch-level objectives \cmt{($\mathcal{L}_{cos^{\beta}}$)} are expressed as:
\begin{equation}
\small
\label{eq:cos+batch}
        \small
        \mathcal{L}_{cos^{+}}=-\mathbb{E}_{j\in[1,B]}\left[\log \frac{\exp(\textit{sim}(z_{s1}^j, z_{s2}^j))}{\sum_{k}^B \exp (\textit{sim}(z_{s1}^j, z_{s2}^k))}\right]
    \end{equation}
    \begin{equation}
    \small
    \label{eq:cos-batch}
    \mathcal{L}_{cos^{-}}=-\mathbb{E}_{j,k\in[1,B]}[\log (1-\textit{sim}(z_{p1}^j, z_{p2}^k))]
    \end{equation}
When a cosine distance objective is used, it is added to the reconstruction objective 
for a final loss term of $\mathcal{L}_{total} = \mathcal{L}_{rec} + \mathcal{L}_{cos}$.

\textbf{Masking}: Drawing on works in masked acoustic token modeling from the generative-based methods\cite{huang2022masked, gong2023contrastive, garcia2023vampnetmusicgenerationmasked}, 
we mask a ratio ($r$) of random entries in the  projected latent subspaces during training. By masking one latent subspace and leaving the other unchanged, the intuition is to encourage the model to rely more on a subset of the subspaces to infer necessary information. We leverage this latent subspace-based masking mechanism as a motivation to enforce the model to learn robust representations that capture the underlying structure in each subspace.


\section{Experimental Design}

\subsection{Dataset and Multi-View Data Pairing}

We use the SONYC-UST-V2 dataset \cite{cartwright2020sonycustv2urbansoundtagging}, which contains $10$-second audio clips recorded from $56$ sensors across New York City as part of the SONYC project \cite{bello2019sonyc}. The 56 sensor classes refer to individual recording devices placed in distinct urban locations, each capturing various city soundscapes\cmt{of various neighborhoods}. Sensors capture important channel effect information, namely the environmental acoustics unique to that location and microphone position\cmt{ (despite being recorded on identical hardware)}. The recordings are non-overlapping in time, meaning that we can obtain multiple audio clips from the same sensor ``for free'' by selecting different temporal segments within a recording\cmt{ from a single sensor}. Therefore, we consider our data pairing strategy as a self-supervised mechanism, as it is similar to sampling pairs of segments from the same recording in the traditional SSL setup. Each audio clip is labeled with one or more of $8$ sound source categories, including engine, alarm, and human voice.




For this study, we use the recording \textit{sensor} as the \textit{shared} factor, pairing clips recorded by the same sensor at different times. 
These randomly selected pairs are assumed to naturally differ in \textit{sound source content}, which we use as the \textit{private} factor in this study.
An example data pair could be two audio recordings from the same sensor location but recorded on different days, where one contains a dog barking sound and another contains an engine and car alarm.




We construct $39k$ training pairs, $6k$ for validation, and $11k$ for testing, containing $39$/$5$/$12$ disjoint sensors respectively. We evaluate on unseen sensors to test the model's generalization ability and robustness for unseen scene variations. For downstream evaluation, we further partition this test set into train, validation, and test subsets stratified by sensor and sound source, resulting in an $8$-class multi-label classification for downstream sound source classification, and $12$-class classification for sensor. Importantly, we only use these labels in downstream evaluation and our core method is fully self-supervised.

\subsection{Audio Preprocessing}
We follow the parameters used for preprocessing audio for the Descript Audio Codec (DAC) \cite{kumar2023highfidelityaudiocompressionimproved}. We resample audio recordings from SONYC to $44.1$KHz and normalize them to $-24$ dB LUFS, following DAC. We pass the full $10$-seconds of audio to the pretrained DAC model\footnote{We use the $44.1$KHz, $8$kbps bitrate pretrained DAC.}. For a $10$-sec. audio clip, this yields an embedding of shape $(862, 1024)$, where $862$ is the number of frames and $1024$ is the feature dimension. 
We use the pre-quantized continuous latents from DAC. Pairs of these embeddings are used as input to our multi-view autoencoder. 


\subsection{Training Recipe}

We train all of our models for $100$ epochs on a single A100 GPU using a batch size of $16$. We use a learning rate of $1e-3$ and AdamW optimization, and perform model selection using minimum total validation loss.




\subsection{Downstream Classification}
We evaluate the informativeness of our learned joint audio representations on downstream sensor and source classification tasks. After training our multi-view autoencoder model, we freeze the trained encoder and use it as a feature extractor to obtain private and shared latents ($z_{p}$ and $z_{s}$) from a single input audio clip to use for downstream training. 
We train independent $2$-layer MLP classifier heads on top of the private, shared, and concatenated latents separately for source and sensor classification. We use cross entropy (CE) as the objective for $12$-class sensor classification, and binary cross entropy (BCE) as the objective for $8$-class source classification in the multi-label setting. 

\begin{table} 
\small
\vspace{0.2cm}
\caption{Downstream source and sensor classification vs. baselines. We present the results of our best-performing model, which uses reconstruction loss and sample-level 
cosine distance loss to separate private latents as the combined training strategy. }

\centering

\begin{tabular}{l|l|c|c}
 &  & \textbf{Source} & \textbf{Sensor} \\
 \textbf{Method}&\textbf{Objective} & \textbf{$n_c=8$} & \textbf{$n_c=12$} \\
\midrule
  \textbf{Multi-view (Best Config.)} & $\mathcal{L}_{rec}$ + $\mathcal{L}_{cos^{-}}$    & $\mathbf{0.633}$ & $\mathbf{0.735}$ \\


  Single-view Autoencoder
  & $\mathcal{L}_{rec}$    & ${0.582}$ & ${0.710}$ \\
  
   Contrastive Learning      & InfoNCE \cite{oord2019representationlearningcontrastivepredictive} & $0.324$ & $0.392$ \\
 \midrule
  DAC \cite{kumar2023highfidelityaudiocompressionimproved}           & N/A    & $0.583$ & $0.684$ \\
 Supervised Learning & BCE/CE & $0.699$ & $0.732$ \\

\end{tabular}
\label{tab: main table}
\end{table}

\subsection{Evaluation Metrics}
\label{evalmetrics}
We evaluate downstream performance using accuracy for 12-class sensor classification, and the Jaccard index\cmt{ (i.e. Intersection over Union)} for 8-class multi-label source classification. For each task, we use the following metrics:

\textbf{Overall accuracy}: We concatenate the private and shared subspaces ($z_{p} \oplus z_{s}$) and use this joint latent as the feature for the downstream task. This gives a measure of overall robustness and information retained in the learned embedding space.

\textbf{Subspace accuracy}: To better-understand the information structure in the latent subspace, 
    we use either $z_{pi}$ or $z_{si}$ as features for downstream classification. This allows us to assess the information present in the separate learned subspaces.
    
\textbf{Directional subspace classification ($DSC\Delta$)}: As a proxy metric to investigate the disentanglement level of the private and shared subspaces, we measure the difference in subspace performance for both source and sensor classification tasks. We define $DSC\Delta$ for the private or shared latent as: 
    
    \begin{equation}
    \small
        \text{DSC}\Delta_{priv} = \text{clf}(z_p)_{\text{source}} - \text{clf}(z_s)_{\text{source}}
    \end{equation}
    \begin{equation}
        \small
        \text{DSC}\Delta_{shared} = \text{clf}(z_s)_{\text{sensor}} - \text{clf}(z_p)_{\text{sensor}},
    \end{equation}
    where $\text{clf}(\cdot)$ indicates the downstream classifier trained per task.
    An ideal scenario for disentanglement in our framework is for the private latent to capture view-specific information (source in this study), and for the shared to capture common information (sensor). 
    If there is no disentanglement and the subspace performance using either latent is identical, this yields $DSC\Delta = 0$.
    Thus, we aim for a positive $DSC\Delta$ score for each latent, indicating the desired disentanglement.






\subsection{Baseline Methods}

We apply the framework above to a series of baseline training strategies, each built upon the pretrained DAC latents as input:

\textbf{DAC \cite{kumar2023highfidelityaudiocompressionimproved} without training}: The naive baseline is training the downstream classifiers directly on top of off-the-shelf DAC latents. This represents the baseline classification potential of the latents before any additional training or disentanglement.

\textbf{Single-view Autoencoder}: We apply the traditional generative learning pipeline using the same encoder and decoder architecture as our method with the reconstruction loss $\mathcal{L}_{rec}$, but with just one view.

\textbf{Contrastive learning}: We apply the traditional contrastive learning framework using the same encoder architecture as our method, but without (1) the notion of separated latent subspaces and (2) using a decoder for reconstruction. This yields one latent space of the same dimensionality as the joint latent in our multi-view based methods.
We train the encoder on pairs of data with 
the same dataset used in our multi-view framework (data paired by common sensor class), utilizing only one training strategy of an InfoNCE \cite{oord2019representationlearningcontrastivepredictive} loss, where each pair is considered a positive sample and samples from different pairs are considered negative samples.

\textbf{Supervised learning}: using the same encoder as our method, with an additional one-layer linear classification head for either multi-label source classification (BCE objective), or multi-class sensor classification (CE objective), without reconstruction. This method provides an upper bound of task performance, not a direct comparison with our proposed methods, since our training is label-free.


\section{Results and Discussion}
We train our model with different combinations of training strategies introduced in Sec. \ref{subsection:designfactor}. Our base method uses a multi-view learning backbone with $\mathcal{L}_{rec}$. Variants of our method are trained using one of the following training strategies on top of the base method: sample or batch-level cosine distance objectives \cmt{($\mathcal{L}_{cos^{S}}$ and $\mathcal{L}_{cos^{\beta}}$ respectively),} applied to the private latents to encourage separation ($\mathcal{L}_{cos^{-}}$) or the shared latents to encourage similarity ($\mathcal{L}_{cos^{-}}$), and lastly masking 
$z_p$ with masking ratio $r\in[0.2, 0.4,0.6,0.8]$. 



\begin{table} 
\small
\caption{Examining the effects of varied objective functions on downstream source and sensor classification.}
\centering
\begin{tabular}{l|l|c|c|c}
\label{tab:2}
 &  & \textbf{Source} & \textbf{Sensor} &  \\
 \textbf{Objective}&\textbf{Configuration} & \textbf{$n_c=8$} & \textbf{$n_c=12$} & \textbf{Avg. Acc.}\\
\midrule
   $\mathcal{L}_{rec}$ & N/A  & $0.610$ & $\mathbf{0.739}$ & $0.675$ \\
    \midrule
     \multirow{2}{*}{$\mathcal{L}_{rec} + \mathcal{L}_{cos-}$} & \textbf{Sample-level}  & $\mathbf{0.633}$ & $0.735$ & $\mathbf{0.684}$ \\
     & Batch-level & $0.610$ & $0.737$  & $0.674$\\
     \midrule
     \multirow{2}{*}{$\mathcal{L}_{rec} + \mathcal{L}_{cos+}$}
    & Sample-level & $0.593$  & $0.719$ & $0.656$\\
    & Batch-level & $0.603$ & $0.728$ & $0.666$\\
    \midrule
    \multirow{1}{*}{$\mathcal{L}_{rec}$ + Mask $z_p$} 
    &  $r=0.4$ & $0.594$ & $0.733$ & $0.664$ 
\end{tabular}
\label{tab:big}
\end{table}

\subsection{Main Comparisons}
\label{sec:comp}
The evaluation results of our best-performing model configurations and baselines are shown in Table \ref{tab: main table}. 
We first observe that the standard contrastive learning approach, trained using data paired by the same sensor class using InfoNCE, yields lower accuracy across the board for both tasks. This indicates that the assumption of commonality in CL is less effective in this nuanced task of sensor classification, and fails at extracting sound source information without an appropriate pairing strategy. Additionally, 
the more traditional single-view autoencoder 
provides 3.9\% relative improvement on sensor classification but no improvement on source classification when compared to DAC latents without any training (denoted with ``N/A'' objective), 
In contrast, even without curating data pairs with matching sound source information, our method combining generative and contrastive principles is able to successfully capture information in both tasks. 

Our best model configuration, using $\mathcal{L}_{rec}$ and sample-level $\mathcal{L}_{cos^{-}}$, improves overall accuracy results by $8.6\%$ and $3.5\%$ for source and sensor classification respectively, compared to the single-view autoencoder baseline. Further, this configuration significantly outperforms the contrastive and DAC-only baselines.
Such improvements indicate the complementary effects of contrastive and generative principles to produce robust environmental sound representations, while overcoming limits of individual traditional SSL methods. 

We also include a fully supervised upper bound for reference. While the supervised sensor classification achieves the highest performance with the help of complete label supervision, our method 
significantly narrows the gap with no explicit prior knowledge about sensor and source information. At the same time, we observe that \cmt{leveraging the same amount of sensor information in }our training framework even leads to a $0.4\%$ relative performance improvement on sensor classification versus a supervised approach.

In Table \ref{tab:2}, we investigate how different objective strategies introduced in Sec. \ref{subsection:designfactor} affect the the learned representations on downstream performance. We find that
separation-based objectives ($\mathcal{L}_{cos-}$) tend to perform better than similarity-based objectives ($\mathcal{L}_{cos+}$), for both sample and batch-level experiments. We do not observe a consistent trend between sample and batch-level cosine distance-based experiments across tasks; for $\mathcal{L}_{cos-}$, sample-level performs better 
(our best configuration in terms of averaged accuracy),
but for $\mathcal{L}_{cos+}$, the batched version is marginally better than sampled. 

\begin{figure}[t]
  \centering 
  \includegraphics[width=0.9\columnwidth, trim={2cm 0.5cm 6cm 1cm}]{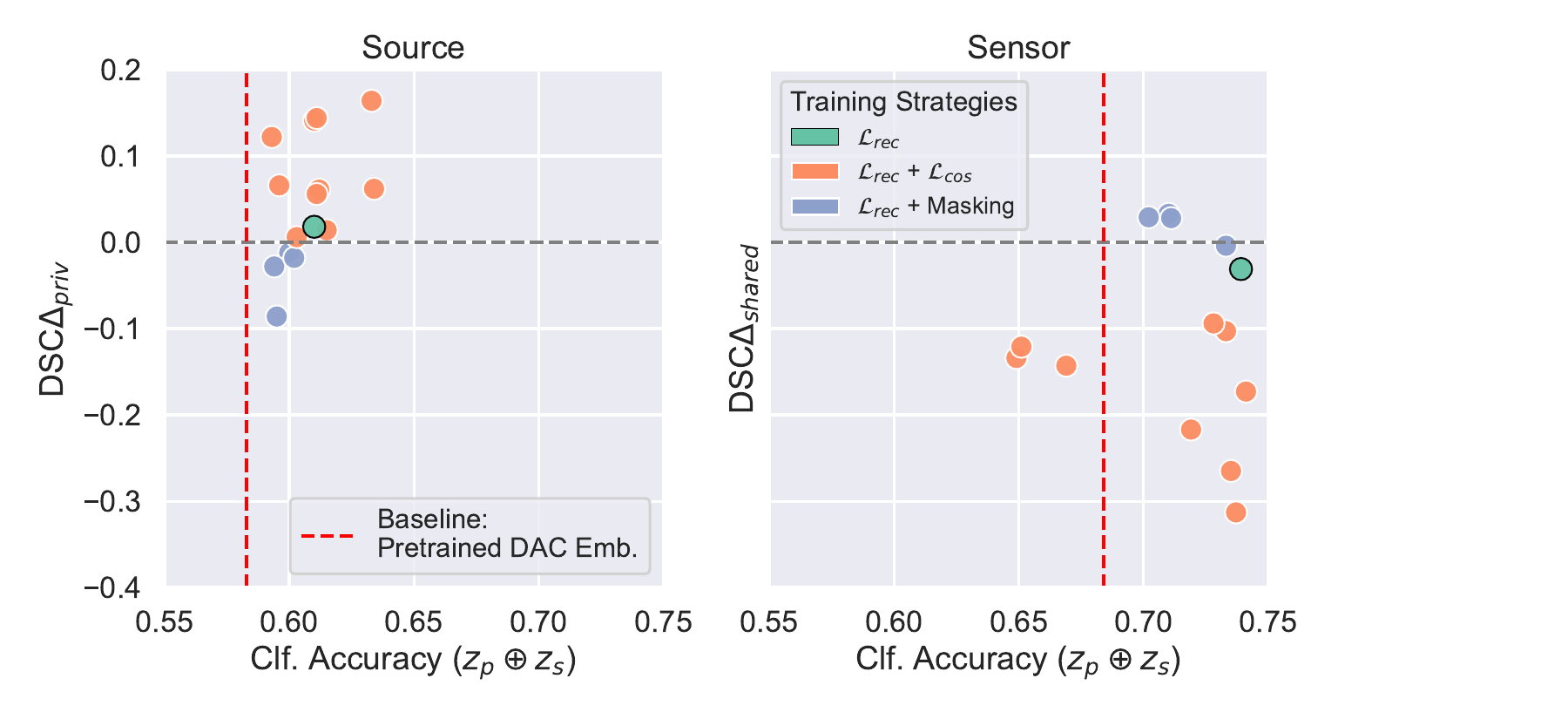}
  \caption{Visualizing the effects of cosine distance and masking strategies on downstream tasks using the overall performance vs. the Directional Subspace Classification metric ($DSC\Delta$). }
  \label{fig:scatter}
\end{figure}

    

 

\subsection{Disentanglement Investigation}
\label{sec:abl}



We also observe the potential for environmental sound attribute disentanglement using our method and investigate this under different training configurations.
In Figure \ref{fig:scatter}, we visualize the trend of information flow for source (left) and sensor (right) attributes grouped by three types of strategies: cosine distance-based disentanglement using only the multi-view backbone with $\mathcal{L}_{rec}$ (green), 
adding $\mathcal{L}_{cos^{+}}$ or $\mathcal{L}_{cos^{-}}$ (orange), or $\mathcal{L}_{rec}$ with masking (blue). 
Each point represents a specific training configuration within one type of training strategy, e.g. a specific masking ratio $r$ within the ``masking'' category, or sample-based $\mathcal{L}_{cos^{+}}$ within  the $\mathcal{L}_{cos}$ grouping. 
The x-axis in Fig. \ref{fig:scatter} represents the overall classification accuracy obtained using the complete latent space ($z_p \oplus z_s$). The red vertical line marks the accuracy scores per task of the DAC baseline as reported in Table \ref{tab: main table}. The y-axis represents the $\text{DSC}\Delta$ to measure the level of information flow in the desired direction as a proxy for disentanglement quality. 

We first show that the majority our multi-view strategies infuse useful source information in the latent subspaces, outperforming the pretrained DAC baseline marked with the red dotted line.
%
We also observe that the multi-view backbone with only $\mathcal{L}_{rec}$ has a positive $DSC\Delta$ for source, but negative $DSC\Delta$ for sensor, indicating that without explicit constraints utilizing multi-view assumptions, the model shows a tendency to steer information about both source and sensor into the private subspace. 
%
While Fig. \ref{fig:scatter} suggests that cosine distance-based strategies generally tend to push both source and sensor information into the private subspace, masking $z_p$ shows the opposite effect; masking shifts overall information into the shared subspace. However, we did not observe obvious trend for different masking ratios. We speculate that the shared sensor information between views may be too subtle to capture in the common subspace with reconstruction alone,
while masking $z_p$ puts more weight on the shared subspace to encode all information to optimize reconstruction.



\section{Conclusion and Future Work}
In this work, we proposed a novel self-supervised multi-view learning framework that integrates contrastive principles within a generative pipeline to improve the robustness of environmental sound representations. 
Our experiments on the SONYC-UST-V2 dataset demonstrate that our method improves downstream performance in both source and sensor classification on recordings in unseen sensors compared to traditional SSL methods. Beyond improved representation robustness, we investigate the effects of different training strategies on latent subspace information flow, showing the potential for environmental sound attribute disentanglement. In the future, we plan to explore using audio understanding models within our framework such as AudioMAE \cite{huang2022masked} or BEATs \cite{beats}, and extend downstream tasks to out-of-distribution data.


\label{section:references}

\section{Acknowledgments}
\label{sec:ack}
We thank our colleagues at Bosch for their support on this work, which is partially supported by award $\#$24-2283 (Robert Bosch, LLC).


\bibliographystyle{IEEEtran}
\bibliography{refs}

\begin{thebibliography}{10}
\providecommand{\url}[1]{#1}
\csname url@samestyle\endcsname
\providecommand{\newblock}{\relax}
\providecommand{\bibinfo}[2]{#2}
\providecommand{\BIBentrySTDinterwordspacing}{\spaceskip=0pt\relax}
\providecommand{\BIBentryALTinterwordstretchfactor}{4}
\providecommand{\BIBentryALTinterwordspacing}{\spaceskip=\fontdimen2\font plus
\BIBentryALTinterwordstretchfactor\fontdimen3\font minus
  \fontdimen4\font\relax}
\providecommand{\BIBforeignlanguage}[2]{{%
\expandafter\ifx\csname l@#1\endcsname\relax
\typeout{** WARNING: IEEEtran.bst: No hyphenation pattern has been}%
\typeout{** loaded for the language `#1'. Using the pattern for}%
\typeout{** the default language instead.}%
\else
\language=\csname l@#1\endcsname
\fi
#2}}
\providecommand{\BIBdecl}{\relax}
\BIBdecl

\bibitem{cola}
A.~Saeed, D.~Grangier, and N.~Zeghidour, ``Contrastive learning of
  general-purpose audio representations,'' in \emph{ICASSP 2021-2021 IEEE
  International Conference on Acoustics, Speech and Signal Processing
  (ICASSP)}.\hskip 1em plus 0.5em minus 0.4em\relax IEEE, 2021, pp. 3875--3879.

\bibitem{byol}
E.~Fonseca, D.~Ortego, K.~McGuinness, N.~E. O’Connor, and X.~Serra,
  ``Unsupervised contrastive learning of sound event representations,'' in
  \emph{ICASSP 2021-2021 IEEE International Conference on Acoustics, Speech and
  Signal Processing (ICASSP)}.\hskip 1em plus 0.5em minus 0.4em\relax IEEE,
  2021, pp. 371--375.

\bibitem{al2021clar}
H.~Al-Tahan and Y.~Mohsenzadeh, ``Clar: Contrastive learning of auditory
  representations,'' in \emph{International Conference on Artificial
  Intelligence and Statistics}.\hskip 1em plus 0.5em minus 0.4em\relax PMLR,
  2021, pp. 2530--2538.

\bibitem{nasiri2021soundclrcontrastivelearningrepresentations}
\BIBentryALTinterwordspacing
A.~Nasiri and J.~Hu, ``Soundclr: Contrastive learning of representations for
  improved environmental sound classification,'' 2021. [Online]. Available:
  \url{https://arxiv.org/abs/2103.01929}
\BIBentrySTDinterwordspacing

\bibitem{MAHYUB2024SLS}
\BIBentryALTinterwordspacing
M.~Mahyub, L.~S. Souza, B.~Batalo, and K.~Fukui, ``Signal latent subspace: A
  new representation for environmental sound classification,'' \emph{Applied
  Acoustics}, vol. 225, p. 110181, 2024. [Online]. Available:
  \url{https://www.sciencedirect.com/science/article/pii/S0003682X24003323}
\BIBentrySTDinterwordspacing

\bibitem{Perera2022adv}
D.~Perera, S.~Essid, and G.~Richard, ``Latent and adversarial data
  augmentations for sound event detection and classification,'' in
  \emph{Proceedings of the 7th Detection and Classification of Acoustic Scenes
  and Events 2022 Workshop (DCASE2022)}, Nancy, France, November 2022.

\bibitem{kumar2023highfidelityaudiocompressionimproved}
\BIBentryALTinterwordspacing
R.~Kumar, P.~Seetharaman, A.~Luebs, I.~Kumar, and K.~Kumar, ``High-fidelity
  audio compression with improved rvqgan,'' 2023. [Online]. Available:
  \url{https://arxiv.org/abs/2306.06546}
\BIBentrySTDinterwordspacing

\bibitem{défossez2022highfidelityneuralaudio}
\BIBentryALTinterwordspacing
A.~Défossez, J.~Copet, G.~Synnaeve, and Y.~Adi, ``High fidelity neural audio
  compression,'' 2022. [Online]. Available:
  \url{https://arxiv.org/abs/2210.13438}
\BIBentrySTDinterwordspacing

\bibitem{gong2022ssast}
Y.~Gong, C.-I. Lai, Y.-A. Chung, and J.~Glass, ``Ssast: Self-supervised audio
  spectrogram transformer,'' in \emph{Proceedings of the AAAI Conference on
  Artificial Intelligence}, vol.~36, no.~10, 2022, pp. 10\,699--10\,709.

\bibitem{chong2023maskedspec}
D.~Chong, H.~Wang, P.~Zhou, and Q.~Zeng, ``Masked spectrogram prediction for
  self-supervised audio pre-training,'' in \emph{ICASSP 2023-2023 IEEE
  International Conference on Acoustics, Speech and Signal Processing
  (ICASSP)}.\hskip 1em plus 0.5em minus 0.4em\relax IEEE, 2023, pp. 1--5.

\bibitem{huang2022masked}
P.-Y. Huang, H.~Xu, J.~Li, A.~Baevski, M.~Auli, W.~Galuba, F.~Metze, and
  C.~Feichtenhofer, ``Masked autoencoders that listen,'' \emph{Advances in
  Neural Information Processing Systems}, vol.~35, pp. 28\,708--28\,720, 2022.

\bibitem{beats}
\BIBentryALTinterwordspacing
S.~Chen, Y.~Wu, C.~Wang, S.~Liu, D.~Tompkins, Z.~Chen, W.~Che, X.~Yu, and
  F.~Wei, ``{BEAT}s: Audio pre-training with acoustic tokenizers,'' in
  \emph{Proceedings of the 40th International Conference on Machine Learning},
  ser. Proceedings of Machine Learning Research, A.~Krause, E.~Brunskill,
  K.~Cho, B.~Engelhardt, S.~Sabato, and J.~Scarlett, Eds., vol. 202.\hskip 1em
  plus 0.5em minus 0.4em\relax PMLR, 23--29 Jul 2023, pp. 5178--5193. [Online].
  Available: \url{https://proceedings.mlr.press/v202/chen23ag.html}
\BIBentrySTDinterwordspacing

\bibitem{abesser2017autoencode}
J.~Abe{\ss}er, S.~I. Mimilakis, R.~Gr{\"a}fe, H.~M. Lukashevich, and
  I.~Fraunhofer, ``Acoustic scene classification by combining autoencoder-based
  dimensionality reduction and convolutional neural networks.'' in
  \emph{DCASE}, 2017, pp. 7--11.

\bibitem{zinemanas2021prototype}
P.~Zinemanas, M.~Rocamora, E.~Fonseca, F.~Font, and X.~Serra, ``Toward
  interpretable polyphonic sound event detection with attention maps based on
  local prototypes.'' in \emph{DCASE}, 2021, pp. 50--54.

\bibitem{Abesser2017dimensionality}
J.~Abeßer, S.~I. Mimilakis, R.~Gräfe, and H.~Lukashevich, ``Acoustic scene
  classification by combining autoencoder-based dimensionality reduction and
  convolutional neural networks,'' in \emph{Proceedings of the Detection and
  Classification of Acoustic Scenes and Events 2017 Workshop (DCASE2017)},
  November 2017, pp. 7--11.

\bibitem{Wilkinghoff2019opensetauto}
K.~Wilkinghoff and F.~Kurth, ``Open-set acoustic scene classification with deep
  convolutional autoencoders,'' in \emph{Proceedings of the Detection and
  Classification of Acoustic Scenes and Events 2019 Workshop (DCASE2019)}, New
  York University, NY, USA, October 2019, pp. 258--262.

\bibitem{xu2021multi}
J.~Xu, Y.~Ren, H.~Tang, X.~Pu, X.~Zhu, M.~Zeng, and L.~He, ``Multi-vae:
  Learning disentangled view-common and view-peculiar visual representations
  for multi-view clustering,'' in \emph{Proceedings of the IEEE/CVF
  international conference on computer vision}, 2021, pp. 9234--9243.

\bibitem{lee2020private}
\BIBentryALTinterwordspacing
M.~Lee and V.~Pavlovic, ``Private-shared disentangled multimodal vae for
  learning of hybrid latent representations,'' 2020. [Online]. Available:
  \url{https://arxiv.org/abs/2012.13024}
\BIBentrySTDinterwordspacing

\bibitem{gong2023contrastive}
\BIBentryALTinterwordspacing
Y.~Gong, A.~Rouditchenko, A.~H. Liu, D.~Harwath, L.~Karlinsky, H.~Kuehne, and
  J.~R. Glass, ``Contrastive audio-visual masked autoencoder,'' in \emph{The
  Eleventh International Conference on Learning Representations}, 2023.
  [Online]. Available: \url{https://openreview.net/forum?id=QPtMRyk5rb}
\BIBentrySTDinterwordspacing

\bibitem{tanakaunsup}
\BIBentryALTinterwordspacing
K.~Tanaka, K.~Yoshii, S.~Dixon, and S.~Morishima, ``Unsupervised
  pitch-timbre-variation disentanglement of monophonic music signals based on
  random perturbation and re-entry training,'' \emph{APSIPA Transactions on
  Signal and Information Processing}, vol.~14, no.~1, pp.~--, 2025. [Online].
  Available: \url{http://dx.doi.org/10.1561/116.20240072}
\BIBentrySTDinterwordspacing

\bibitem{luo2020unsupervised}
Y.-J. Luo, K.~W. Cheuk, T.~Nakano, M.~Goto, and D.~Herremans, ``Unsupervised
  disentanglement of pitch and timbre for isolated musical instrument sounds.''
  in \emph{ISMIR}, 2020, pp. 700--707.

\bibitem{wilkins2024selfsupervisedmultiviewlearningdisentangled}
\BIBentryALTinterwordspacing
J.~Wilkins, S.~Ding, M.~Fuentes, and J.~P. Bello, ``Self-supervised multi-view
  learning for disentangled music audio representations,'' 2024. [Online].
  Available: \url{https://arxiv.org/abs/2411.02711}
\BIBentrySTDinterwordspacing

\bibitem{bello2019sonyc}
J.~P. Bello, C.~Silva, O.~Nov, R.~L. Dubois, A.~Arora, J.~Salamon, C.~Mydlarz,
  and H.~Doraiswamy, ``Sonyc: A system for monitoring, analyzing, and
  mitigating urban noise pollution,'' \emph{Communications of the ACM},
  vol.~62, no.~2, pp. 68--77, 2019.

\bibitem{wu2022wav2clip}
H.-H. Wu, P.~Seetharaman, K.~Kumar, and J.~P. Bello, ``Wav2clip: Learning
  robust audio representations from clip,'' in \emph{ICASSP 2022-2022 IEEE
  International Conference on Acoustics, Speech and Signal Processing
  (ICASSP)}.\hskip 1em plus 0.5em minus 0.4em\relax IEEE, 2022, pp. 4563--4567.

\bibitem{radford2021learning}
A.~Radford, J.~W. Kim, C.~Hallacy, A.~Ramesh, G.~Goh, S.~Agarwal, G.~Sastry,
  A.~Askell, P.~Mishkin, J.~Clark \emph{et~al.}, ``Learning transferable visual
  models from natural language supervision,'' in \emph{International conference
  on machine learning}.\hskip 1em plus 0.5em minus 0.4em\relax PMLR, 2021, pp.
  8748--8763.

\bibitem{guinot2024leaveoneequivariantalleviatinginvariancerelatedinformation}
\BIBentryALTinterwordspacing
J.~Guinot, E.~Quinton, and G.~Fazekas, ``Leave-one-equivariant: Alleviating
  invariance-related information loss in contrastive music representations,''
  2024. [Online]. Available: \url{https://arxiv.org/abs/2412.18955}
\BIBentrySTDinterwordspacing

\bibitem{oord2019representationlearningcontrastivepredictive}
\BIBentryALTinterwordspacing
A.~van~den Oord, Y.~Li, and O.~Vinyals, ``Representation learning with
  contrastive predictive coding,'' 2019. [Online]. Available:
  \url{https://arxiv.org/abs/1807.03748}
\BIBentrySTDinterwordspacing

\bibitem{garcia2023vampnetmusicgenerationmasked}
\BIBentryALTinterwordspacing
H.~F. Garcia, P.~Seetharaman, R.~Kumar, and B.~Pardo, ``Vampnet: Music
  generation via masked acoustic token modeling,'' 2023. [Online]. Available:
  \url{https://arxiv.org/abs/2307.04686}
\BIBentrySTDinterwordspacing

\bibitem{cartwright2020sonycustv2urbansoundtagging}
\BIBentryALTinterwordspacing
M.~Cartwright, J.~Cramer, A.~E.~M. Mendez, Y.~Wang, H.-H. Wu, V.~Lostanlen,
  M.~Fuentes, G.~Dove, C.~Mydlarz, J.~Salamon, O.~Nov, and J.~P. Bello,
  ``Sonyc-ust-v2: An urban sound tagging dataset with spatiotemporal context,''
  2020. [Online]. Available: \url{https://arxiv.org/abs/2009.05188}
\BIBentrySTDinterwordspacing

\end{thebibliography}

%
%
%
%
%
%
%
%
%

\end{document}